\documentclass[fleqn,usenatbib]{mnras}

\usepackage{newtxtext,newtxmath}

\usepackage[T1]{fontenc}

\DeclareRobustCommand{\VAN}[3]{#2}
\let\VANthebibliography\thebibliography
\def\thebibliography{\DeclareRobustCommand{\VAN}[3]{##3}\VANthebibliography}

\usepackage{graphicx}	
\usepackage{amsmath}	
\usepackage{multirow}
\usepackage{xcolor}

\newcommand{\Teff}{$T_{\text{eff}}$}
\newcommand{\logg}{$\log(g)$}
\newcommand{\loggunits}{$\log(g/\text{cm s$^{-2}$})$}
\newcommand{\drvm}{$\Delta \text{RV}_{\text{max}}$}
\newcommand{\drvpp}{$\Delta \text{RV}_{\text{pp}}$}
\newcommand{\kms}{km s$^{-1}$}
\newcommand{\vsini}{$v\sin{i}$}
\newcommand{\vsinir}{$v\sin(i_{\text{rot}})$}
\newcommand{\msun}{M$_{\odot}$}
\newcommand{\porb}{P$_{\text{orb}}$}
\newcommand{\prot}{P$_{\text{rot}}$}
\newcommand{\pcrit}{P$_{\text{crit}}$}
\newcommand{\iorb}{$i_{\text{orb}}$}
\newcommand{\irot}{$i_{\text{rot}}$}
\newcommand{\agesd}{$\tau_{\text{SD}}$}
\newcommand{\agegyro}{$\tau_{\text{gyro}}$}

\title[Stellar Multiplicity and Rotation]{Stellar multiplicity and stellar rotation: Insights from APOGEE}

\author[C. M. Daher et al.]{
Christine Mazzola Daher,$^{1}$\thanks{E-mail: c.mazzola.daher@pitt.edu}
Carles Badenes,$^{1}$
Jamie Tayar,$^{2,\dagger}$
Marc Pinsonneault,$^{3}$
\newauthor
Sergey E. Koposov,$^{4,5}$
Kaitlin Kratter,$^{6}$
Maxwell Moe,$^{6}$
Borja Anguiano,$^{7}$
\newauthor
Diego Godoy-Rivera,$^{3}$
Steven Majewski,$^{7}$
Joleen K. Carlberg,$^{8}$ 
Matthew G. Walker,$^{9}$
\newauthor
Rachel Buttry,$^{9}$
Don Dixon,$^{10}$
Javier Serna,$^{11}$
Keivan G. Stassun,$^{10}$
\newauthor
Nathan De Lee,$^{12}$
Jes\'us Hern\'andez,$^{11}$
Christian Nitschelm,$^{13}$
Guy S. Stringfellow,$^{14}$
\newauthor
and Nicholas W. Troup$^{15}$
\\
$^{1}$Department of Physics and Astronomy and Pittsburgh Particle Physics, Astrophysics and Cosmology Center (PITT PACC), University of Pittsburgh,\\3941 O‘Hara Street, Pittsburgh, PA 15260, USA\\
$^{2}$Institute for Astronomy, University of Hawaii, Honolulu, HI 96822, USA\\
$^{\dagger}$Hubble Fellow\\
$^{3}$Department of Astronomy, The Ohio State University, Columbus, OH 43210, USA\\
$^{4}$ Institute for Astronomy, University of Edinburgh, Royal Observatory, Blackford Hill, Edinburgh EH9 3HJ, UK\\
$^{5}$ Institute of Astronomy, University of Cambridge, Madingley Road, CB3 0HA, UK\\
$^{6}$Steward Observatory, University of Arizona, 933 North Cherry Avenue, Tucson, AZ 85721, USA\\
$^{7}$Department of Astronomy, University of Virginia, Charlottesville, VA, 22904, USA\\
$^{8}$Space Telescope Science Institute, 3700 San Martin Drive, Baltimore, MD 21218, USA\\
$^{9}$McWilliams Center for Cosmology, Department of Physics, Carnegie Mellon University, 5000 Forbes Avenue, Pittsburgh, PA 15213, USA\\
$^{10}$Department of Physics \& Astronomy, Vanderbilt University, Nashville, TN 37235, USA\\
$^{11}$Instituto de Astronom\'{i}a, Universidad Aut\'{o}noma de M\'{e}xico, Ensenada, B.C, M\'{e}xico\\
$^{12}$Department of Physics, Geology, and Engineering Tech, Northern Kentucky University, Highland Heights, KY 41099, USA\\
$^{13}$Centro de Astronom{\'i}a (CITEVA), Universidad de Antofagasta, Avenida Angamos 601, Antofagasta 1270300, Chile\\
$^{14}$Center for Astrophysics and Space Astronomy, Department of Astrophysical and Planetary Sciences, University of Colorado, Boulder, CO 80309, USA\\
$^{15}$Department of Physics, Salisbury University, Salisbury, MD 21801, USA\\
}

\date{Accepted XXX. Received YYY; in original form ZZZ}

\pubyear{2021} 

\begin{document}
\label{firstpage}
\pagerange{\pageref{firstpage}--\pageref{lastpage}}
\maketitle

\begin{abstract}
We measure rotational broadening in spectra taken by the Apache Point Observatory Galactic Evolution Experiment (APOGEE) survey to characterise the relationship between stellar multiplicity and rotation. We create a sample of 2786 giants and 24 496 dwarfs with stellar parameters and multiple radial velocities from the APOGEE pipeline, projected rotation speeds \vsini\ determined from our own pipeline, and distances, masses, and ages measured by Sanders \& Das. We use the statistical distribution of the maximum shift in the radial velocities, \drvm, as a proxy for the close binary fraction to explore the interplay between stellar evolution, rotation, and multiplicity. Assuming that the minimum orbital period allowed is the critical period for Roche Lobe overflow and rotational synchronization, we calculate theoretical upper limits on expected \vsini\ and \drvm\ values. These expectations agree with the positive correlation between the maximum \drvm\ and \vsini\ values observed in our sample as a function of \logg. We find that the fast rotators in our sample have a high occurrence of short-period ($\log(P/\text{d})\lesssim 4$) companions. We also find that old, rapidly-rotating main sequence stars have larger completeness-corrected close binary fractions than their younger peers. Furthermore, rapidly-rotating stars with large \drvm\ consistently show differences of 1-10 Gyr between the predicted gyrochronological and measured isochronal ages. These results point towards a link between rapid rotation and close binarity through tidal interactions. We conclude that stellar rotation is strongly correlated with stellar multiplicity in the field, and caution should be taken in the application of gyrochronology relations to cool stars.
\end{abstract}

\begin{keywords}
binaries: close -- binaries: spectroscopic -- stars: rotation -- stars: evolution
\end{keywords}



\section{Introduction}
\label{sec:intro}

Stellar multiplicity plays a crucial role in stellar astrophysics, with roughly half of the solar-type stars in the solar neighbourhood being part of multi-star systems \citep{Duequennoy1991,Raghavan2010}. In cases where these stars get close enough to interact, they can be responsible for a whole host of astrophysical phenomena, ranging from low- and high-mass X-ray binaries, Type Ia SNe, many core-collapse SNe, novae, cataclysmic variables, and the majority of stellar sources of gravitational waves \citep[see][for a review]{Marco2017}. Most of these phenomena arise from the interplay between stellar evolution and multiplicity. Data from astrometric surveys now makes it possible to study stellar multiplicity across the Hertzsprung-Russell diagram, and as a function of many stellar properties  \citep{Belokurov2020,Price-Whelan2020,Mazzola2020}, though spectroscopy remains the best way to identify and characterize the close ($\log(P/\text{d})\lesssim 4$), unresolved binaries that are the progenitors of interacting systems. Luckily, modern spectroscopic surveys have already collected the necessary data for large samples of stars within the Milky Way.

One such survey is the Apache Point Observatory Galactic Evolution Experiment 2 \citep[APOGEE-2;][]{Majewski2017}, a component of the Sloan Digital Sky Survey IV \citep[SDSS-IV;][]{Gunn2006, Blanton2017}. APOGEE-2 has two high-resolution ($R\sim22 500$), multiplexed infrared spectrographs \citep{Wilson2019}, deployed in the northern and southern hemispheres, that have taken multi-epoch data for hundreds of thousands of stars representative of every major component in our Galaxy. Spectral parameters are determined via the the APOGEE Stellar Parameter and Chemical Abundances Pipeline \citep[ASPCAP;][]{Perez2016,Joensson2020}, and precise radial velocities (RVs) are calculated for each individual visit spectra \citep[][]{Nidever2015}. Though most APOGEE-2 stars only have sparsely-sampled RV curves, a large number of such curves can effectively constrain stellar multiplicity statistics and their correlation with a number of spectroscopic parameters \citep{Badenes2012,Maoz2012,Sana2012}.

\begin{figure}
	\includegraphics[width=\columnwidth]{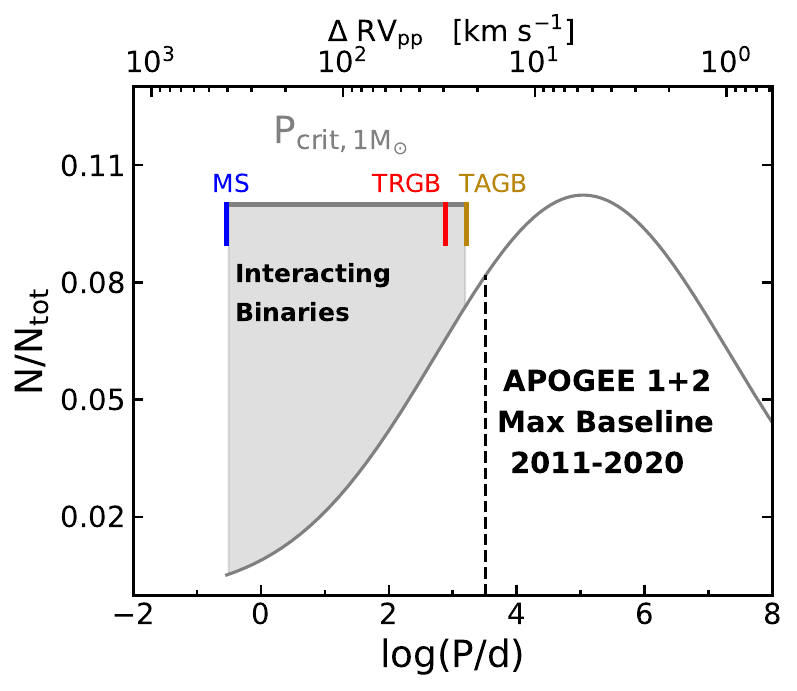}
    \caption{The \citet{Raghavan2010} lognormal period distribution ($\overline{\log{\text{P}}}=5.04$, $\sigma_{\log{\text{P}}}=2.28$) for Sun-like MS stars in the solar neighbourhood. Values of \pcrit\ (equation~\ref{eq:pcrit}) are indicated for a 1 \msun\, [Fe/H] $=0$ star in an equal-mass ($q=1$) binary across several important evolutionary points. The top axis shows $\Delta\text{RV}_{\text{pp}}$ (equation~\ref{eq:drvpp}) for a 1 \msun\ star in an $e=0$, \iorb$=90^{\circ}$ binary across the range of periods shown.}
    \label{fig:rag}
\end{figure}

\cite{Badenes2018} and \cite{Price-Whelan2020} found a strong relationship between the close binary fraction and the surface gravities measured by APOGEE, \loggunits, which is an observational proxy for the evolutionary stage of stars. Both studies observed a positive correlation between the close binary fraction and \logg, and measured a low fraction of close binaries for red clump (RC) stars. These results agree with the expectation that companion engulfment occurs at longer and longer periods as the primary ascends the red giant branch (RGB), resulting in the gradual attrition of short-period companions until stars reach the tip of the red giant branch (TRGB). The end product of this process is a small fraction (a factor of $\sim$2 fewer) of close binaries for core-He burning stars in the RC. The shortest possible orbital period for a given system should correspond to the critical period for Roche Lobe overflow (RLOF),
\begin{equation}
    \text{P}_{\text{crit}} = \frac{2\pi}{\sqrt{\mathcal{R}^{3}(q)(1+q)}} \left( \frac{GM}{g^{3}} \right)^{1/4}
    \label{eq:pcrit}
\end{equation}
where $g$ is the surface gravity and $M$ the mass of the primary, $q=M_{2}/M$ is the system's mass ratio, and $\mathcal{R}(q)$ is the ratio between the radius of the Roche Lobe and orbital separation \citep{Eggleton1983}. Values for \pcrit\ for several evolutionary stages are shown in Fig.~\ref{fig:rag}, together with with the lognormal period distribution measured by \cite{Raghavan2010} for Sun-like dwarfs in the Solar neighborhood. The top axis shows the maximum observable RV shift for a given system: its peak-to-peak RV amplitude, which is twice the semi-amplitude $K$ and given by
\begin{equation}
    \Delta\text{RV}_{\text{pp}} = 2K = \frac{2}{\sqrt{1-e^{2}}} \left( \frac{\pi GM}{2\text{P}_{\text{orb}}} \right)^{1/3} \sin(i_{\text{orb}})
	\label{eq:drvpp}
\end{equation}
where $e$ is the eccentricity, $M$ is the primary's mass, \porb\ is the orbital period of the system, and \iorb\ is the inclination angle of the orbital axis.

\begin{figure}
	\includegraphics[width=\columnwidth]{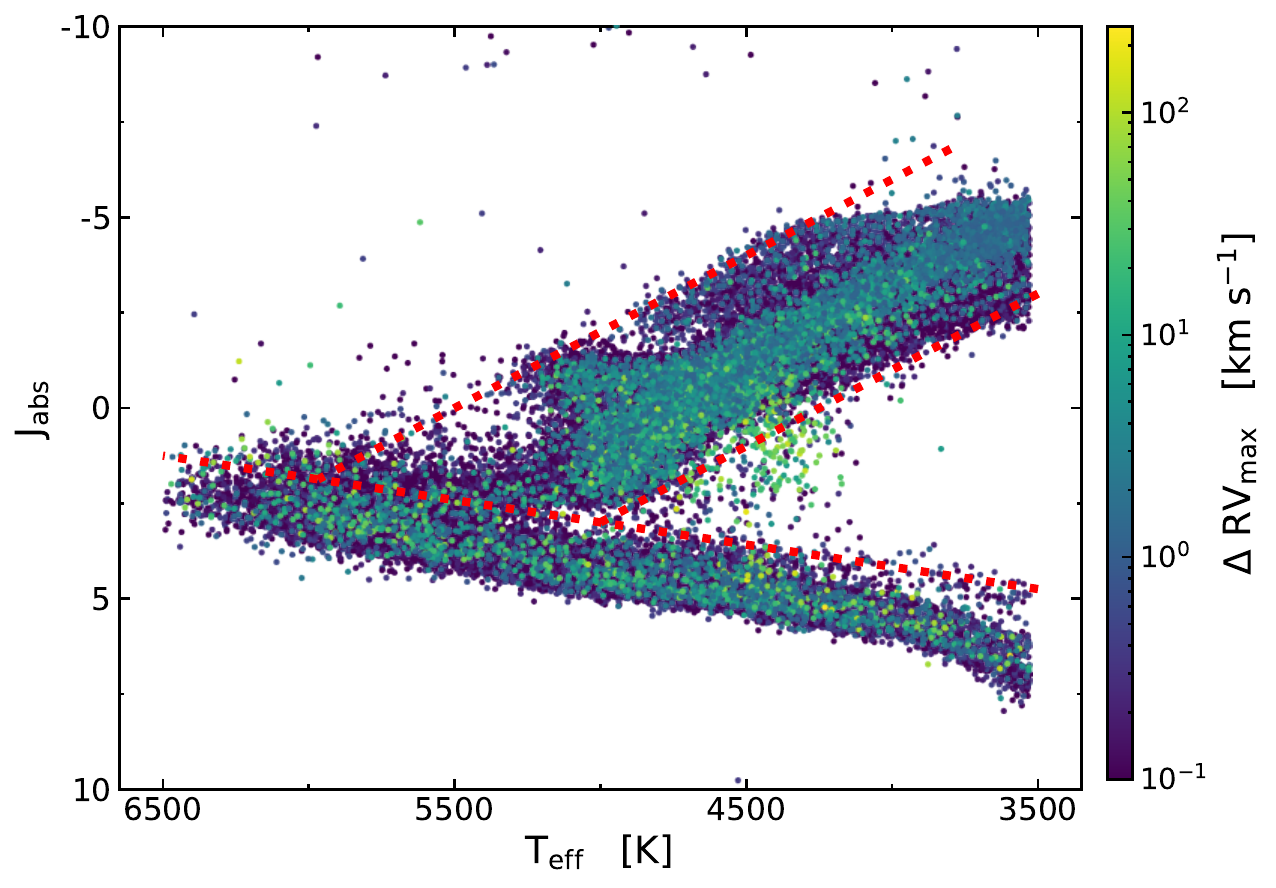}
    \caption{An HR diagram for stars passing our quality cuts using APOGEE DR14 uncalibrated \Teff\ and absolute 2MASS $J$ magnitude, calculated using the \citet{Sanders2018} distance estimates. Points are coloured by the \drvm\ colorbar at right. Stars with \drvm$>1$\kms\ are shown on top for clarity and are present across the entire range of the HR diagram, meaning we are sensitive to binaries across all evolutionary points and metallicities. The dotted red lines indicate the regions we use to define our ``red giant'' and ``dwarf'' samples. These cuts retain the sequence of photometric binaries seen above the primary main sequence.}
    \label{fig:hrdiag}
\end{figure}

\begin{figure*}
	\includegraphics[width=0.8\textwidth]{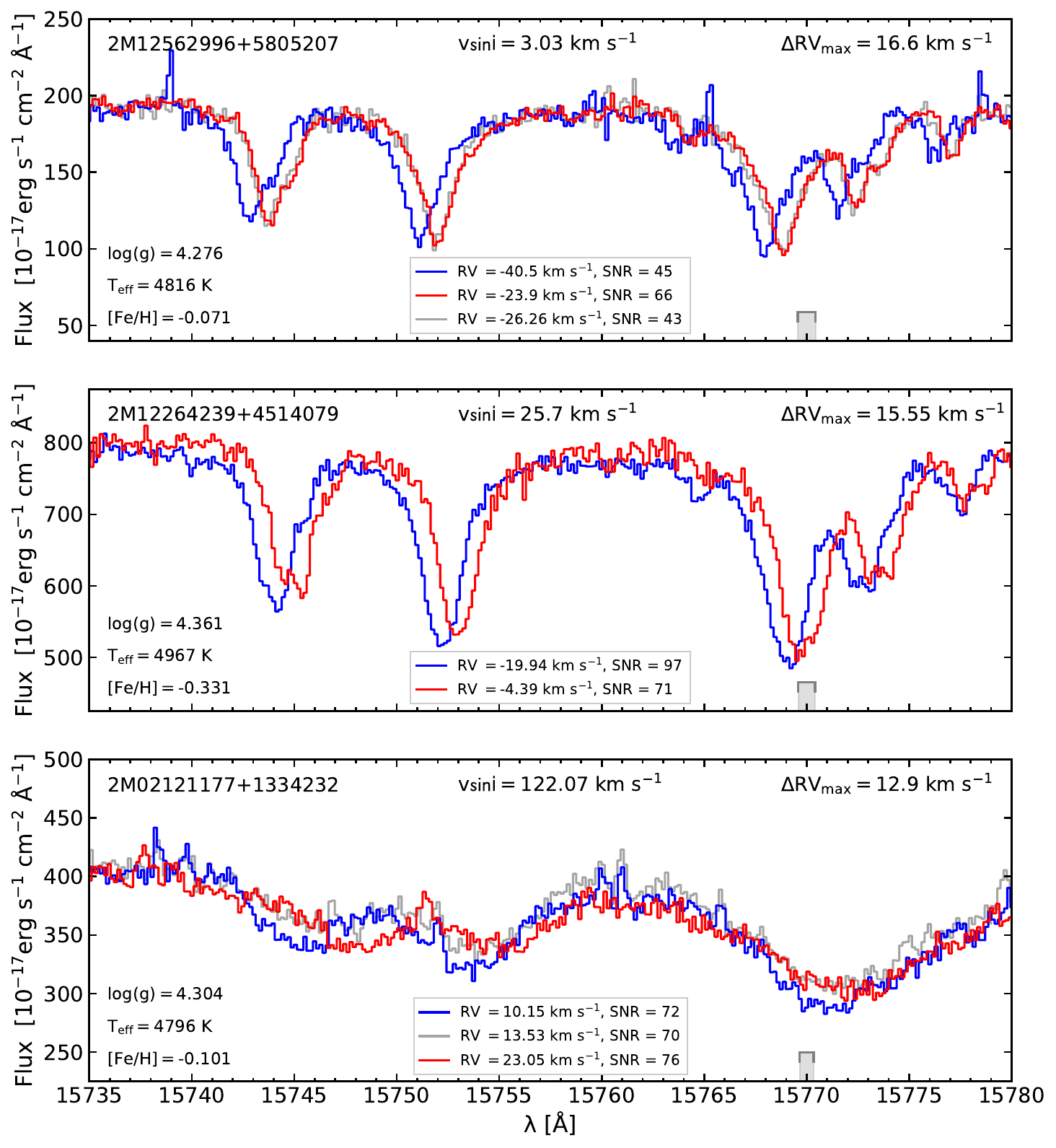}
    \caption{Visit spectra for three dwarfs with similar stellar parameters (\logg, \Teff, metallicity) and \drvm\ but a range of \vsini\ values representative of our sample. The top panel shows a slowly-rotating dwarf, the middle panel a progressively faster rotator, and the bottom panel an extremely fast rotator relative to our sample. The visit with maximum RV is shown in red and the minimum RV shown in blue, and the three major absorption lines are all from magnesium. The grey ruler at the bottom right of each panel shows the expected shift in wavelength from each star's \drvm\ for a feature centered on $\lambda=15770$ \AA. Even with the extremely broadened lines in the final panel, we can still identify RV shifts from APOGEE spectra.}
    \label{fig:samp_spectra}
\end{figure*}

Short-period systems can interact in a variety of ways before mass transfer or companion engulfment occurs. Tidal dissipation in close binaries will lead to rotational synchronization,  equalizing the orbital period and the rotation periods, as well as orbital circularization, though the mechanisms for dissipation differ between early and late type stars \citep[see][]{Mazeh2008, Zahn2008}. There is a clear ``circularization period'' in close binaries, below which orbital eccentricities are usually very small \citep[see][and references therein]{Price-Whelan2018a}, and which varies with \logg. Rotational synchronization timescales are thought to be two or three orders of magnitude smaller than orbital circularization timescales, but depend on eccentricity \citep{Mazeh2008}. Between the onset of rotational synchronization and the final circularization of the orbit, multiple systems can experience ``pseudosynchronization'' where the star's rotation speed is synchronized to its orbital speed at periastron \citep{Mazeh2008, Zahn2008}, but full synchronization can be further delayed or prevented due to three-body effects \citep{Lurie2017}. The timescale for orbital circularization depends strongly on the orbital period and the structure of the star's convective and radiative envelopes \citep{Verbunt1995}. Together, these effects produce pitfalls for gyrochronology, the method of calculating a star's age based upon a decrease in its rotation speed as it ages. Due to rotational synchronization, stars in close binaries may have rotation speeds much greater than their single counterparts of the same age \citep[e.g.][]{Simonian2019}.

The goal of this paper is to characterize the dependence of stellar rotation on close binary interactions and evolutionary state via reasonable assumptions about rotational synchronization using public data from APOGEE and \textit{Gaia}, stellar parameters measured from this data, and rotation speeds derived from rotational line broadening in the APOGEE spectra \citep{Tayar2015, Dixon2020}. So far, the connection between intrinsic stellar parameters, stellar multiplicity, and stellar rotation has been explored mainly using small samples in stellar clusters \citep[for a review, see][]{Mazeh2008}, or more recently with rotation periods from \textit{Kepler} \citep{Simonian2019,Simonian2020}. Here, we aim to provide a broad study of these relationships using a large sample of field stars, with a specific focus on the implications for gyrochronology. We discuss our sample selection and pipeline in Section~\ref{subsec:sampleselec} and our theoretical framework in Section~\ref{subsec:theory}, present our results in Sections~\ref{sec:results} and \ref{sec:gyro}, and draw our conclusions in Section~\ref{sec:concl}.

\section{Data and Methods}
\label{sec:datamethods}

\subsection{Sample Selection}
\label{subsec:sampleselec}
Spectral parameters are taken from the APOGEE Data Release 14 \verb+allStar+ file, which contains entries for 277 371 objects \citep{Abolfathi2018, Holtzman2018, Joensson2018}. Note that only 258 475 unique APOGEE IDs exist among these entries; this is because the same star may be observed on different fibre plugplates that each correspond to a different field centre, but ASPCAP does not automatically combine all visit spectra with the same APOGEE ID from different fields. However, the combination of an APOGEE ID and field location ID can securely identify each unique target, its combined spectrum, its associated stellar parameters in the \verb+allStar+ file, and its individual visit RVs in the \verb+allVisit+ file.

Our first round of quality cuts removed stars with the STAR\_BAD flag set in the ASPCAP bitmask \citep{Holtzman_2015}. Stars identified as commissioning observations \citep[bit 1 in STARFLAG][]{Holtzman_2015} and telluric calibrators \citep[bit 9 in both the apogee target2 and apogee2 target2 masks][]{Zasowski2013,Zasowski2017} were removed as well. We limited our sample to field stars by removing known cluster members (bit 9 in apogee target1 and apogee2 target1 and bit 10 in apogee target2 and apogee2 target2).

We required that stars have well measured ($\neq-9999$, the default for a bad value), uncalibrated effective temperatures (\Teff) and surface gravities [\logg], because most APOGEE DR14 dwarfs do not have calibrated \logg\ values \citep{Holtzman2018}. In this work, we chose to restrict our view to giants and dwarfs, so we first removed all stars identified as RC in the APOGEE DR14 RC catalog \citep{Bovy2014}. We then estimated deredenned $JHK_{s}$ magnitudes from the $A_{k}$ value adopted in targeting \citep[AK\_TARG,][]{Zasowski2013,Zasowski2017} and used them to further remove potential RC stars based upon the criteria outlined in \cite{Price-Jones2017}.

Using the VISITS\_PK indices \citep{Holtzman_2015, Nidever2015}, we identified the individual visits that are included in the combined APOGEE spectrum for each APOGEE ID/location ID combination and required that two or more of the visit spectra had S/N$\geq40$. We concatenated all acceptable visit RVs for stars with duplicated APOGEE IDs, meaning stars with at least one good visit in two or more fields are included in our data set. For these objects, any stellar parameters with multiple valid ($\neq-9999$) values were averaged. We do not use the RV uncertainties in quality cuts and instead use the \drvm\ distributions to inform our estimates for the RV precision in Section~\ref{sec:results}.

Close companions with sufficiently high $q$ can produce measurable contributions to the observed spectrum, which can introduce biases into APOGEE's spectral fits and resulting spectral parameters \citep{El-Badry2018a}. Additionally, these double-lined spectroscopic binaries, or SB2s, can lead to two distinct peaks in the spectrum's cross-correlation function and thus unreliable identification of the primary's RVs. We account for this uncertainty in the RV selection by using the sample of likely SB2s presented in \cite{Mazzola2020}, where the RV for each visit was determined at the highest peak of either the APOGEE CCFs or recalculated CCFs using the method described in \citet{Kounkel2019,Kounkel2021}. Our quality-cut sample contained 108 789 stars, 1052 of which were identified as likely SB2s.

\begin{figure}
	\includegraphics[width=\columnwidth]{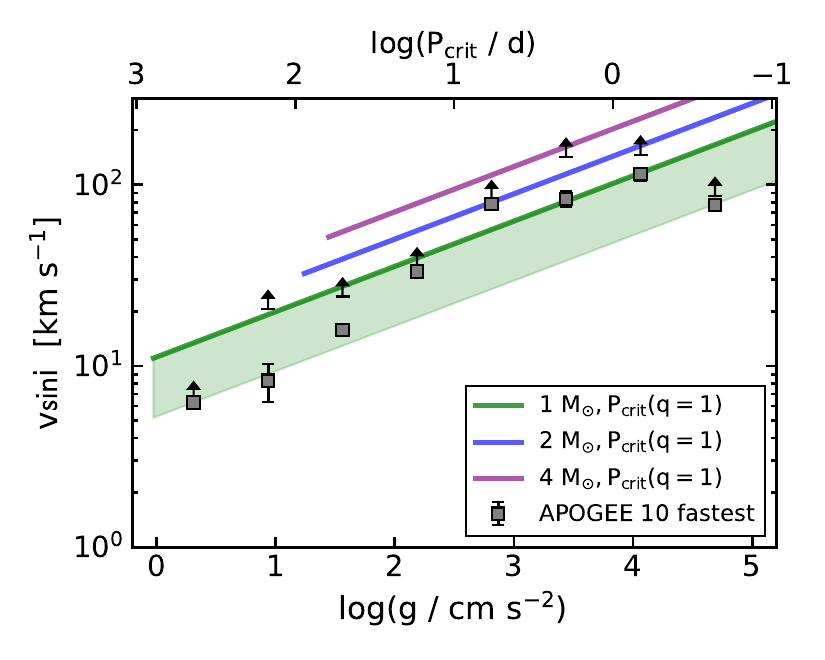}
    \caption{Relationship between \vsini\ and \logg\ accompanied by theoretical upper limits. The median \vsini\ for the ten fastest rotators in each \logg\ sample are shown as grey squares with error bars indicating Poisson uncertainties, and black arrows indicate the largest \vsini\ in each bin. The diagonal lines are theoretical constraints for several masses assuming $q=1$, perpendicular rotational axes ($i_{\text{rot}}=90^{\circ}$), and tidal synchronization at the beginning of RLOF (\porb $=$ \prot $=$ \pcrit). Because our sample is biased against high-$q$ systems after removing suspected SB2s, the green shaded region indicates $0.25\leq q \leq1.0$. The top axis shows \pcrit\ (equation~\ref{eq:pcrit}) for a binary with $q=1$, $M=1$ \msun\ across the range of \logg\ values. The colored lines end at the TRGB in MIST models of solar metallicity.}
    \label{fig:loggmax}
\end{figure}

The APOGEE ASPCAP pipeline estimates rotational broadening for most dwarfs but nearly no giants or subgiants. Following the work of \citet{Tayar2015,Dixon2020}, we cross-correlate the observed combined spectrum with broadened versions of the provided fit spectrum to determine if additional broadening is necessary over what was identified in the ASPCAP analysis. Because the APOGEE spectra are recorded on three individual detector arrays, this procedure was performed on the data from each of the arrays separately, giving us a mean and standard deviation for each additional broadening measurement. We excluded measurements below 5 \kms\ for giants, allowing measurements down to 2 \kms\ for dwarfs, and excluded any cases where the standard deviation was larger than the average. We then added this additional broadening to any \vsini\ measurement provided by the ASPCAP pipeline; for some stars where there was no \vsini\ provided before, we add new measurements. 

After running our quality-cut APOGEE sample through this pipeline, we then removed the likely SB2s within our sample due to concerns that their rotation rates are likely overestimated due to our pipeline interpreting blended spectral lines as rotational broadening \citep{Simonian2020}. This will bias our sample of binaries towards mass ratios $q\lesssim0.9$, and this may introduce a slight bias towards higher primary masses, as there is some evidence that K-type stars have a larger twin excess than FG stars \citep{ElBadry2019}. However, an individual analysis of each SB2 spectrum and estimated \vsini\ would be required to confidently include them, and such an analysis is beyond the scope of the current work.

We used APOGEE DR14 rather than the newest public (DR16) or proprietary (DR17) data so that we could cross-match our sample against the \cite{Sanders2018} catalogue to add estimates of mass $M$, distance $d$, and age $\tau$. \cite{Sanders2018} provides Bayesian posteriors for these parameters by using broadband photometry, spectral parameters from a number of spectroscopic surveys, including APOGEE DR14, and \textit{Gaia} DR2 parallaxes to fit PARSEC isochrones. Requiring non-NAN values for $d$, $M$, and $\tau$ resulted in a final sample with 104 987 stars. The HR diagram for our final sample is shown in Fig.~\ref{fig:hrdiag}, with absolute 2MASS magnitude $J_{\text{abs}}$ plotted versus APOGEE uncalibrated \Teff. The points are coloured by the maximum RV shift, \drvm$=|$RV$_{\text{max}}-$RV$_{\text{min}}|$ \citep[see also][]{Badenes2012, Maoz2012, Badenes2018, Moe2019, Mazzola2020}. Many of the same features noted in \cite{Mazzola2020} are also clear in this similar sample, e.g. high \drvm\ stars spread across the entire HR diagram, though we note that the sequence of high-\drvm\ stars in the photometric binary sequence is far less prounounced without the SB2s that were included in that prior work. For further commentary, we refer the reader to the discussion in that work. Using the distribution from these parameters, we make one final cut to select giants and dwarfs with the red dotted lines shown in Fig.~\ref{fig:hrdiag}, yielding a sample of 79 308 giants and 24 768 dwarfs. The lines are drawn such that we remove outliers from the bulk of the distribution while retaining the offset photometric binary sequence among the dwarfs at the expense of also retaining some subgiants, which we will revisit in Section~\ref{sec:gyro}. These cuts may also remove potential blue stragglers, but as these stars may have anomalously large rotation speeds as a result of binary mass transfer or mergers \citep[see][and discussion within]{Leiner2019}, a deeper analysis would be required to accurately interpret any observations and is beyond the scope of the current work.

Further requiring a measured value for \vsini\ limits our sample to 2786 giants and 24 496 dwarfs. Below is a summary of the differences between the ASPCAP measurements for these stars and the values we use in this work: 
\begin{enumerate}
    \item 1851 giants and 23 865 dwarfs had no changes to their ASPCAP \vsini\ measurements;
    \item 46 giants and 516 dwarfs had additional rotational broadening between $0< \Delta v \sin{i} < 10$ \kms;
    \item 40 giants and 98 dwarfs had added rotational broadening of $\Delta v\sin{i} \geq10$ \kms;
    \item 849 giants and 17 dwarfs did not have ASPCAP \vsini\ values and so our pipeline provides measurements.
\end{enumerate}
Fig.~\ref{fig:samp_spectra} shows visit spectra for several representative stars with large \drvm, similar \logg, \Teff, and [Fe/H], and a range of \vsini\ values. Though the star in the bottom panel demonstrates extreme rotational broadening relative to the other stars, the high-quality spectra from APOGEE are still able to confidently distinguish large variation in stellar RVs.

\begin{figure}
	\includegraphics[width=\columnwidth]{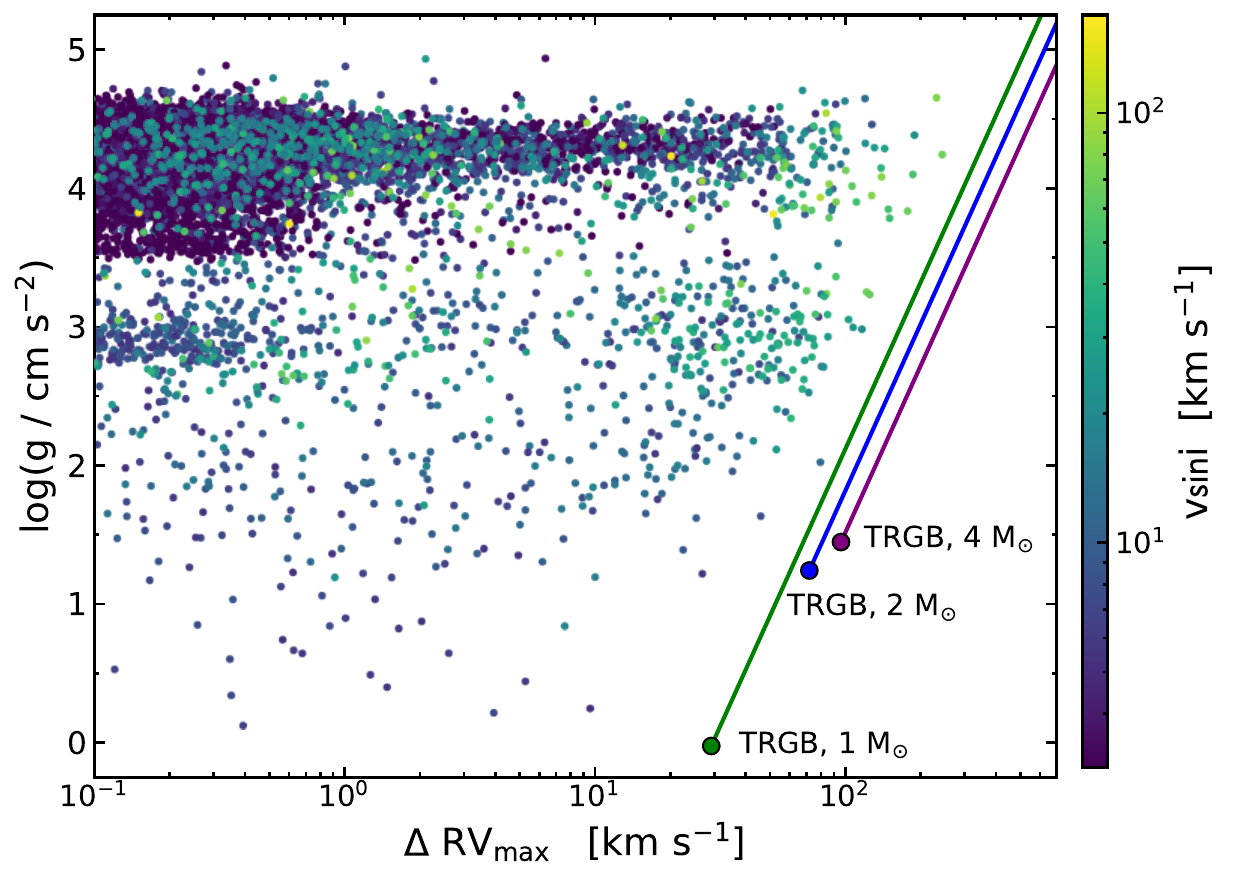}
    \caption{Distribution of \drvm\ and \logg\ for our sample with a colour bar on \vsini\, which shows a correlation between large \drvm\ and \vsini. The diagonal lines show the maximum \drvm\ values expected as a function of \logg\ at 1 (green), 2 (blue), and 4 \msun\ (purple), which are calculated from equation~(\ref{eq:drvpp}) assuming $q=1$, $e=0$, \iorb$=90^{\circ}$, and \porb$=$\pcrit. Terminal symbols show the position of the TRGB.}
    \label{fig:logg_drvm_vsinicb}
\end{figure}

\subsection{Theoretical Framework}
\label{subsec:theory}

In the simplest approximation, we expect the orbital properties of a short-period binary to dictate the rotation period of its constituent stars through spin-orbit synchronization. In that case, the properties of the orbit predict the range of allowed stellar rotation velocities, and we can use this allowed range to interpret the distributions of measured values and relate them to the physics of tidal interactions in binary systems.

Assuming a star is spherical and lacks surface differential rotation, we can define its rotation speed as
\begin{equation}
    v\sin(i_{\text{rot}}) = \frac{2\pi R}{\text{P}_{\text{rot}}} \sin(i_{\text{rot}}) = \frac{2\pi}{\text{P}_{\text{rot}}} \sqrt{\frac{GM}{g}}\sin(i_{\text{rot}})
    \label{eq:vsini}
\end{equation}
where $g$ is the surface gravity, \prot\ is the period of rotation, and \irot\ is the inclination angle of the axis of rotation. This ignores surface differential rotation, which has been observed with \textit{Kepler} in single stars \citep{Reinhold2013, Reinhold2015} and more recently in eclipsing binaries \citep{Lurie2017,Jermyn2020}. However, Zeeman Doppler imaging indicates that surface differential rotation in fast rotating stars is very small, so for the sake of calculating upper limits, we will ignore differential rotation and assume we can relate the observed rotation speed \vsinir\ to a singular rotational period \prot\ via equation~(\ref{eq:vsini}).

 We distinguish between the orbital and rotational inclination angles because it still unclear what the preferred system alignment is \citep{Justesen2020}. Neither of these angles are directly measurable from our data, and so we will refer to our measured rotation speeds as \vsini\, where $i$ includes effects from both rotational and orbital inclination angles.

\begin{figure}
	\includegraphics[width=\columnwidth]{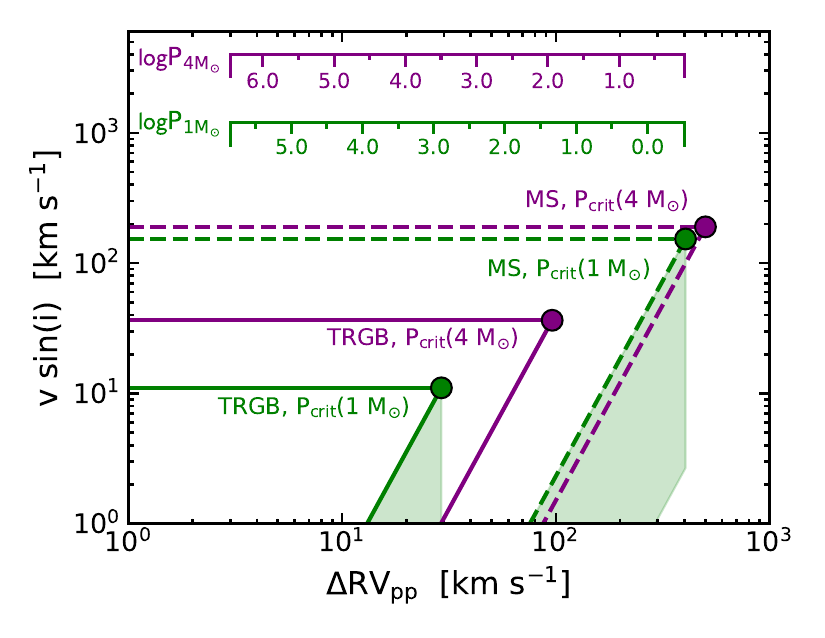}
    \caption{Theoretical constraints on $\Delta\text{RV}_{\text{pp}}$ and \vsini\ assuming $q=1$, edge-on ($i_{\text{orb}}=90^{\circ}$), circular orbits, perpendicular rotational axes ($i_{\text{rot}}=90^{\circ}$), and tidal synchronization (\porb $=$ \prot). The diagonal lines are calculated using equations~{\ref{eq:drvpp}-\ref{eq:vsini}} for P $\geq$ \pcrit, where the circular point shows \pcrit\ and hence the start of RLOF (equation~\ref{eq:pcrit}). The \logg\ values for each point are taken from MIST models at solar metallicity. The MS points are \loggunits\ $=4.546$ (1 \msun) and $4.316$ (4 \msun), while the TRGB are \loggunits\ $=-0.024$ (1 \msun) and $1.447$ (4 \msun). Shaded regions indicate the range expected for $1^{\circ} \leq i_{\text{rot}}\leq 90^{\circ}$; for each point on the diagonal, follow the shaded region straight down to see the range of expected \vsini\ values. Horizontal lines emphasise the maximum \vsini\ expected at the start of RLOF. The coloured rulers indicate the $\log(\text{P}/\text{d})$, where P$=$\porb $=$ \prot, that corresponds to the \drvpp\ values along the x-axis for each mass.}
    \label{fig:drvm_rot_theory}
\end{figure}

\begin{figure*}
	\includegraphics[width=\textwidth]{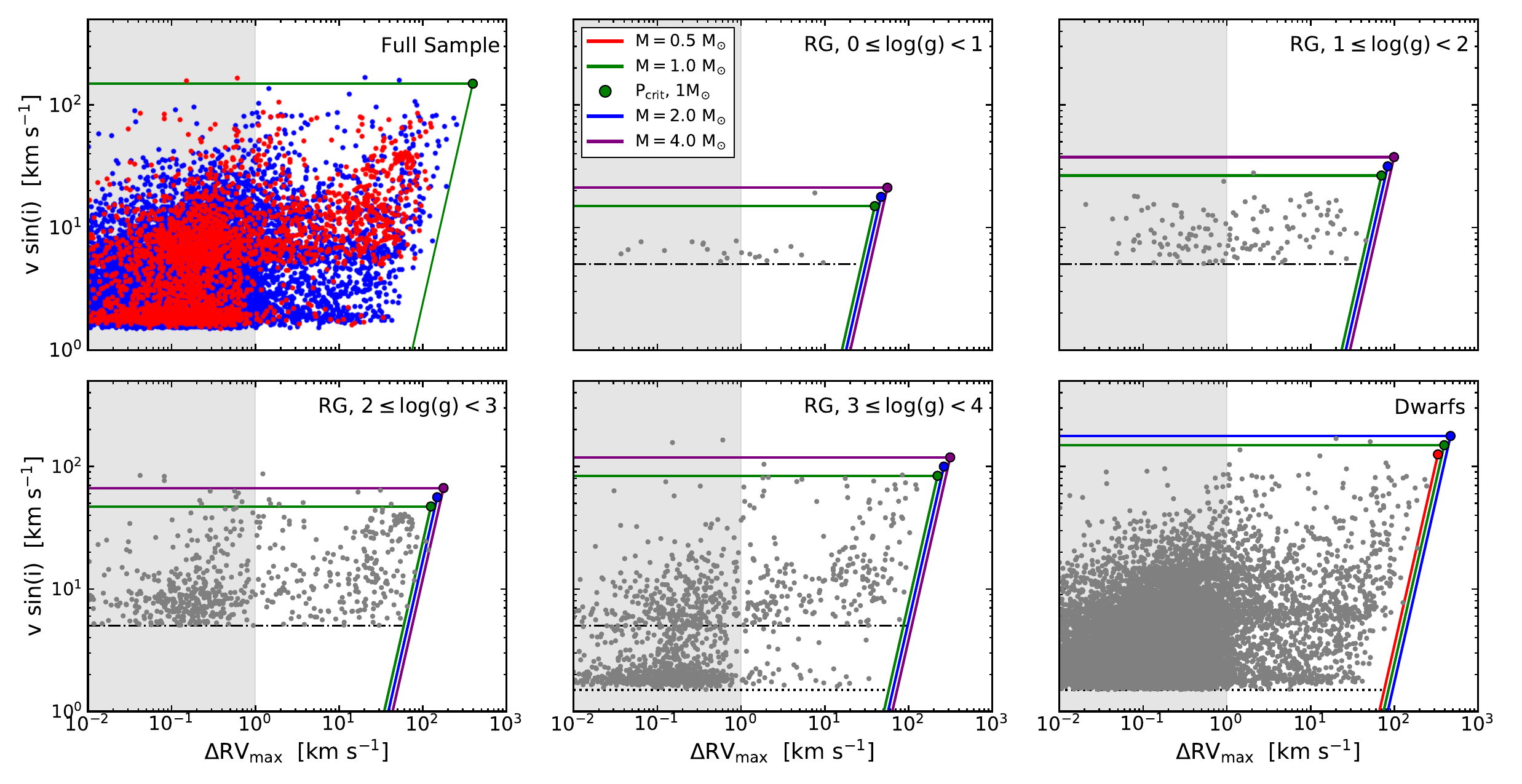}
    \caption{Distribution of \vsini\ and \drvm\ plotted alongside theoretical constraints assuming orbital synchronization. The diagonal lines show the same theoretical constraints shown in Fig.~\ref{fig:drvm_rot_theory} calculated at the midpoint of the \logg\ bin and a range of masses representative of the \citet{Sanders2018} mass distribution for that \logg\ sample. We distinguish two regions of interest, likely RV variables from stars with noise-dominated RV variation, by hatching \drvm\ $\leq1$ \kms. The dash-dot line shows the minimum \vsini\ measurement from our pipeline, $v\sin{i}=5$ \kms, and the dotted line in the final two panels show the approximate minimum value from ASPCAP, $v\sin{i}=1.5$ \kms. Top left: the red and blue points designate our giants and dwarfs samples, respectively. Remaining panels: grey points indicate the main sample across various \logg\ bins.}
    \label{fig:drvm_rot_data}
\end{figure*}

We can relate our understanding of stellar evolution to expectations for rotational synchronization by looking at the relationship between \logg\ and \vsini. As a Sun-like star ascends the RGB, its \logg\ will decrease and its radius will increase until it reaches the TRGB. This gradually reduces the allowed range of semi-major axis for a detached nearby companion as \pcrit\ for RLOF increases, the tidal period lengthens, and the maximum \vsini\ decreases. In this framework, we can calculate theoretical upper limits for measured quantities by assuming stellar masses that are representative of our sample and combining equations~({\ref{eq:pcrit}) and (\ref{eq:vsini}}) under these assumptions:
\begin{enumerate}
    \item equal mass binary, $q=1$
    \item rotation axis perpendicular to line-of-sight, $i_{\text{rot}}=90^{\circ}$
    \item \pcrit\ is the minimum possible period
    \item orbital synchronization occurring at the minimum period, \prot$=$\porb$=$\pcrit
\end{enumerate}
The resulting relationships between \vsini\ and \logg\ are shown as the solid lines in Fig.~\ref{fig:loggmax}. We use solar metallicity models from the MESA Isochrone and Stellar Tracks collaobration \citep[MIST;][]{Paxton2011,Paxton2013,Paxton2015,Choi2016,Dotter_2016} to determine the location of the TRGB (corresponding to the maximum radius and lowest \logg) for each mass. Removing the suspected SB2s has biased our sample against high-$q$ systems, so the green shaded region shows the range of \vsini\ expected for $0.25\leq q \leq 1.0$. We compare these theoretical limits to the median \vsini\ of the ten fastest rotators in each \logg\ bin, shown as the grey squares with error bars showing Poisson uncertainties, and the maximum observed \vsini\ for each bin, indicated by the black arrows. The maximum observed rotation speed for each bin is strongly driven by the stars' \logg, and the simple set of assumptions laid out above are able to reproduce the general trend exhibited by the APOGEE data.

\section{Results: The relationship between rotation, evolution, and stellar multiplicity}
\label{sec:results}

We now consider the joint relationship between RV variability, stellar evolution, and stellar rotation through \drvm, \logg, and \vsini\ in Fig.~\ref{fig:logg_drvm_vsinicb}. Starting from assumptions listed previously, we further require an edge-on ($i_{\text{orb}}=90^{\circ}$), circular ($e=0$) orbit to find the maximum expected \drvpp\ values as a function of \logg. These constraints are shown as the diagonal lines, and the maximum \drvm\ values in our sample are nicely bounded by them, as \citet{Badenes2018} noted with APOGEE DR13 data. Stars with \drvm\ $\lesssim1$ \kms\ may be the result of RV uncertainties and should not be treated as a true detection of RV variability. As we discuss in more detail later, true RV variables can begin to be identified from \drvm\ $\gtrsim1$ \kms; in our sample, 627 giants (23 per cent) and 1556 dwarfs (6.4 per cent) have \drvm\ $>1$ \kms. The points in Fig.~\ref{fig:logg_drvm_vsinicb} are coloured by \vsini, and we observe a colour gradient with \drvm: stars with large RV variability are more likely to have large \vsini\ values. More quantitatively, we can compare the ratio of rapid rotators (\vsini\ $\geq 10$ \kms) to slow rotators (\vsini\ $<10$ \kms) for RV variables and non-RV variables. Giants with \drvm\ $\geq10$ \kms\ have $N_{\text{fast}}/N_{\text{slow}}=223/84=2.65$, whereas giants with \drvm\ $<10$ \kms\ have $N_{\text{fast}}/N_{\text{slow}}=434/2045=0.212$; for the dwarfs, these fractions are $178/293=0.608$ and $1254/22771=0.055$, respectively.

To explore this relationship further, we relax our previous orbital synchronization condition (\prot$=$\porb$=$\pcrit) and allow the synchronized period to be any value that is equal to or larger than the critical period for RLOF (\prot$=$\porb$\geq$\pcrit). We will use these assumptions to produce upper limits on \drvpp\ and \vsini\ and directly compare with our observed \drvm\ and \vsini. Before we compare with data, we display these constraints for two different masses and two \logg\ values corresponding to solar-type main sequence (MS) and TRGB evolutionary phases in Fig.~\ref{fig:drvm_rot_theory}. The shaded regions show the range of expected rotation speeds if we vary \irot\ from $1-90^{\circ}$. The horizontal lines provide an expected upper limit on the rotation speed for binaries with a given mass and \logg. Note that this is not a hard upper limit, like the break-up velocity, but is simply the point where RLOF begins and the system will experience more complex interactions due to mass transfer.

Fig.~\ref{fig:drvm_rot_data} displays these constraints against the measurements for our APOGEE dwarfs and giants. The upper left panel shows the giants in red and the dwarfs in blue, together with the theoretical upper limits calculated for a star with $M=1$ \msun, \logg$=4.5$. The dwarfs are shown by themselves in the final panel, and the remaining panels show the giants split into four bins in \logg.

The theoretical upper limits prove generally successful at constraining the data in both axes, and we recover the predicted upper limit on \vsini\ as a function of \logg.  From theory and Fig.~\ref{fig:loggmax}, we expect that stars with higher \logg\ values are able to rotate faster due to their smaller radii, and combined with our characterisation of RLOF via \pcrit, the horizontal lines constrain the vast majority of the observed \vsini\ values. The few exceptions in the bottom panels are discussed in more detail below.

The hatched region from \drvm\ $\leq1$ \kms\ is intended to guide the eye in distinguishing stars whose RV variability is most likely driven by RV uncertainties from \textit{bona fide} RV variables that are likely to have short-period companions. For an extended discussion of the influence of stellar properties and RV errors on the distribution of \drvm, we refer the reader to \citet{Badenes2018}.

For now, we will make a few observations about general trends in these groups of stars. 
\begin{enumerate}
    \item In every panel with adequately large samples, stars with \drvm\ $\geq1$ \kms\ appear to show \vsini\ values that span the entire range up to the theoretical limit.
    \item In the lower panels, stars with \drvm\ $<1$ \kms\ (hatched region) demonstrate two roughly distinct populations. There is a noisy trend of decreasing rotation speed with decreasing \drvm; as \vsini\ decreases, spectral lines become sharper, making RV measurements more accurate. While we still caution against inferring binarity from RV variability at these low \drvm, it is worth noting the overall trend between rotation and \drvm. However, there are some systems with \vsini\ that extend up to or just above the horizontal lines, and these are discussed in more detail in point (v).
    \item The dot-dashed horizontal lines (\vsini\ $=5$ \kms) and dotted lines (\vsini\ $=1.5$ \kms) indicate the minimum \vsini\ measurement from our pipeline or ASPCAP. We removed giants with \vsini\ $<5$ \kms\ from our sample due to the pipeline's unreliability below this threshold \citep{Tayar2015}. The pipeline is more reliable for dwarfs, who have narrower lines in general due to their larger \logg, so we allowed dwarfs to have \vsini\ down to the minimum measurement value. However, the RG bin with $3\leq$ \logg\ $< 4$ contains a mixture of giants and subgiants and thus a mixture in the lower allowed \vsini\ measurement.
    \item From Fig.~\ref{fig:drvm_rot_theory}, a star with a large \drvm\ value can have an unfavourable \irot\ and thus a \vsini\ value that would be beyond the diagonal line. A few stars lie along the boundaries, particularly in the dwarf and RG $2\leq$ \logg\ $< 3$ panels, but none lie significantly beyond the diagonal lines. Measuring rotation through \vsini\ necessarily biases our sample against low \irot, so our data is unable to distinguish whether this lack of result is due to selection effects or to something more fundamental, such as preferential spin-orbit alignment.
    \item The vast majority $(>85$ per cent) of the stars that lie off the sequence described in (ii), including all six stars with low \drvm\ and rotation speeds above the horizontal lines, have sparsely sampled RV curves with only two or three visits that pass our quality cuts, so these may be true binaries whose visits are unfortunately timed in the orbit. Out of the six stars above the upper limits, the minimum and maximum RVs for one star (2M18423451-0422454) are separated by roughly seven months, but the other five outliers have all their visits spaced out over two to four days. Additional RVs could help distinguish whether these are unluckily sampled binaries or single stars with extremely fast rotation and will be the subject of future investigations.
\end{enumerate}

Estimates for the orbital periods would provide great insight into the extent orbital synchronization explains our high-\drvm\ and high-\vsini\ systems. We cross-matched our stars against the \textit{Gold Sample} from \cite{Price-Whelan2020} to add estimates for orbital parameters from their custom Monte Carlo sampler, \textit{The Joker}. This resulted in 45 dwarfs and 19 giants, which was too small of a sample for us to draw meaningful conclusions. These results are unsurprising, since tightly constraining orbital parameters from an RV curve typically requires a large number of RVs \citep[for a discussion, see][]{Price-Whelan2020}, but the majority of the stars in our sample have only two (41.4 per cent) or three (43.4 per cent) visits. Expanding the number of stars in our sample with estimated orbital parameters is the subject of future work.

\begin{figure}
	\includegraphics[width=\columnwidth]{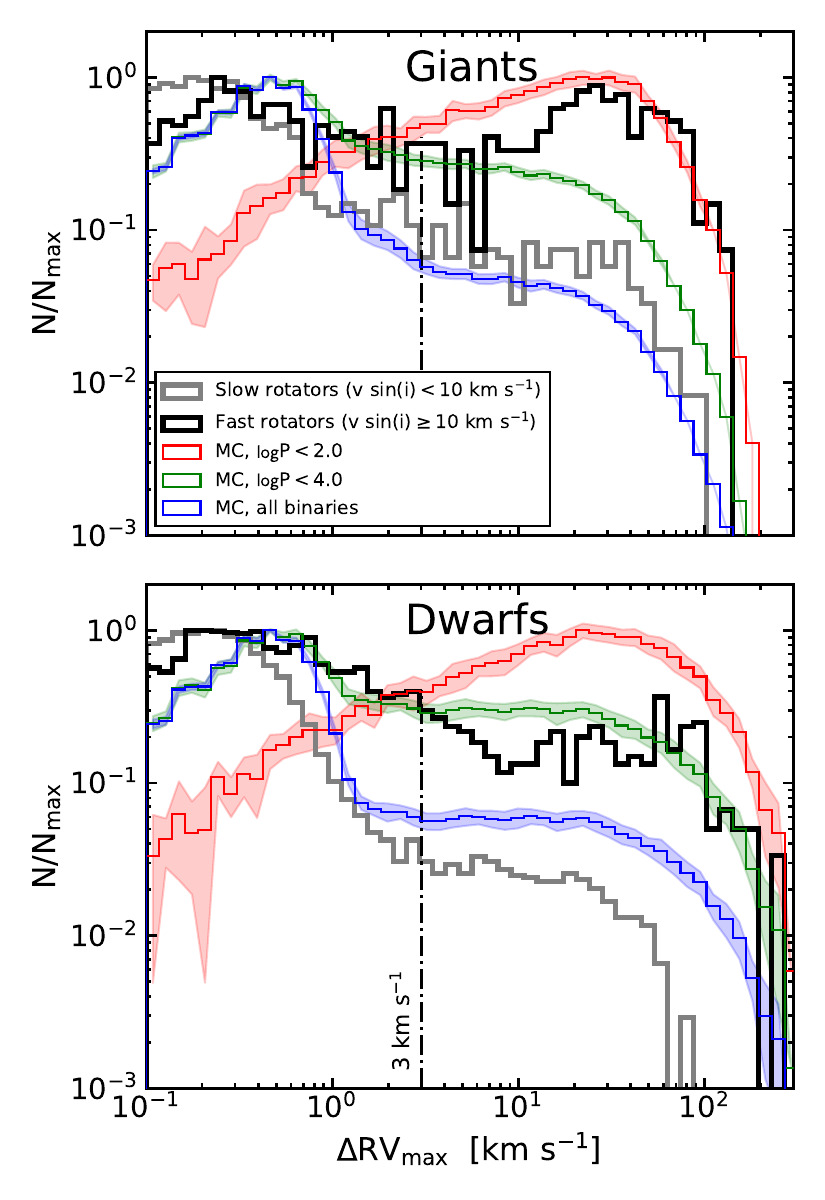}
    \caption{Normalized \drvm\ histograms with slow and fast rotators shown in grey and black, respectively. The red, green, and blue lines are for several $\log(\text{P}/\text{d})$ samples from the relevant MC simulations, with shading indicating $1 \sigma$ intervals from bootstrapping the sample ($N_{boots}=50$). The tail of RV variables appears to begin around \drvm{} $=3$ \kms\ for the slow rotators but is closer to $10$ \kms\ for the rapid rotators. The distributions for the rapid rotators in both dwarfs and giants display a broad core (\drvm{} $\lesssim 10$ \kms), likely connected to larger RV uncertainties from rotationally broadened lines, and a prominent tail (\drvm{} $\gtrsim 10$ \kms) that, upon comparison with the MC simulations, suggests these samples contain a significant fraction of short period binaries.}
    \label{fig:drvm_hists}
\end{figure}

Fig.~\ref{fig:drvm_hists} compares the normalized \drvm\ histograms for our slow and rapid rotators (\vsini\ $<10$ \kms\ and \vsini\ $\geq10$ \kms) in grey and black, respectively. The histogram bin values are normalized by the largest histogram bin, which is $N^{\text{slow}}_{\text{max}}=1369$ and $N^{\text{fast}}_{\text{max}}=60$ for the dwarfs and $N^{\text{slow}}_{\text{max}}=121$ and $N^{\text{fast}}_{\text{max}}=27$ for the giants. The slow rotators display the standard features of a high-quality statistical sample of sparsely sampled RV curves: a core of low \drvm\ values dominated by RV uncertainties, and a tail of genuine RV variables dominated by short-period binaries \citep{Maoz2012}. A vertical line at \drvm$=3$ \kms\ shows the clear transition from core to tail seen in dwarf slow rotators, though this transition from core to tail is less clear for the giants because larger RV uncertainties, on average, contribute to a broader core \citep{Badenes2018}. The distributions for the rapid rotators deviate considerably from this model, with a much broader core that extends out to \drvm\ $\lesssim10$ \kms\ and then an equivalent fraction of RV variable systems with \drvm\ $\gtrsim10$ \kms. The rapid rotators' broader core is due to larger RV uncertainties from their rotationally broadened lines, as Fig.~\ref{fig:samp_spectra} demonstrates. With these caveats in mind, we will use the threshold \drvm\ $\geq 3$ \kms\ as an indicator for RV variability in Section~\ref{sec:gyro} but will include a comparison to the threshold \drvm\ $\geq 10$ \kms\ in the text.

To help interpret these distributions, we rely on the Monte Carlo (MC) sampler described in \citet{Mazzola2020}, which is built to simulate populations of mock multiple systems and predict their \drvm. We defer a detailed discussion of the MC sampler to those in \cite{Badenes2018} and \citet{Mazzola2020} and instead briefly describe the MC and list the settings used in this work. Each simulated binary is assigned its main orbital parameters from observational distributions and randomly assigned an orbital inclination and initial phase. We then simulate observations of the system's RVs using the randomly assigned visit history of a real APOGEE star in our sample, with RV errors drawn from a user-specified distribution. In this work, primary masses are drawn from the \cite{Sanders2018} mass distributions, and we assume a flat mass ratio distribution between $0.1\leq q \leq 0.9$ \citep{Moe2017}, allowing no binaries with $q>0.9$ to account for our sample's bias against equal-mass binaries from the removal of likely SB2s. The remaining settings are listed in Table~\ref{tab:MC_params}; the number of stars in each mock sample, $N$, was chosen to be ten times the the number of objects in the corresponding \logg\ bin from our data to allow for bootstrapping uncertainty regions but keeping the relative fraction of systems the same. The simulations corresponding to the first four rows of Table~\ref{tab:MC_params} were combined when looking at the APOGEE giants as a whole (like in Fig.~\ref{fig:drvm_hists}), but could be separately analysed if desired. The simulation corresponding to the fifth row was used to compare against our APOGEE dwarfs.

The red, green, and blue histograms in Fig.~\ref{fig:drvm_hists} show normalized \drvm\ distributions for various samples of MC-simulated binaries. The blue histogram shows all of our simulated binaries, which have a clear core and tail similar to the slow rotators in APOGEE. The green histogram corresponds to all close ($\log(\text{P}/\text{d})<4.0$) binaries; here, too, there is a significant fraction of systems with low \drvm, but the tail of RV variables is much larger. The red histogram represents the closest binaries ($\log(\text{P}/\text{d})<2.0$), which are completely dominated by RV variables. It also has very few stars at low \drvm\ because the large \drvpp{} values make it highly unlikely to observe the primary at points in its orbit with similar RV, though not impossible, as suggested by the few stars with high \vsini\, low \drvm\, and three or fewer visits discussed previously. For both dwarfs and giants, the \drvm\ distributions of the fast rotators in APOGEE show prominent tails of RV variables, indicating that these samples have a large fraction of short period binaries.

\begin{table*}
    \centering
    \caption{Parameters used to produce mock data from MC simulations.}
    \label{tab:MC_params}
    \begin{tabular}{cccccc}
        \hline
        $N$    & multiplicity fraction $f_{m}$              & \loggunits & RV error source     & RV error $\mu$  (\kms)       & RV error $\sigma$  (\kms) \\
        \hline
        30 000   & \multirow{5}{*}{0.5} & 0.5                  & \multirow{5}{*}{Gaussian} & \multirow{5}{*}{0.0} & 0.75              \\
        230 000 &                      & 1.5                  &                           &                      & 0.25              \\
        340 000 &                      & 2.5                  &                           &                      & 0.25              \\
        200 000 &                      & 3.5                  &                           &                      & 0.25              \\
        260 000 &                      & 4.5                  &                           &                      & 0.25              \\
        \hline
    \end{tabular}
\end{table*}

\section{Discussion: Implications for Gyrochronology}
\label{sec:gyro}

Observations of open clusters and star forming regions have revealed that protostars are born with a wide range of rotation rates \citep{Kraft1965,Vogel1981,Stauffer1987}. Regardless of their initial rotation speed, stars with masses below $1.3$ \msun, corresponding to \Teff\ $<6250$ K, tend to spin down over their MS lifetimes, whereas higher-mass stars maintain the rotation they were imparted at their births much more effectively \citep{Wolff1986}. This mass threshold is known as the Kraft break \citep{Kraft1967} and is thought to be due to the presence, or lack thereof, of convective envelopes \citep{Durney1978}. Stars below the Kraft break have convective envelopes which drive magnetized winds that cause angular momentum loss, slowing the star's rotation rate as it ages. Gyrochronology relations use the empirical constraints from clusters \citep{Gallet2015,GodoyRivera2021} and field stars \citep{Angus2015,Angus2020} to parametrise this angular momentum loss and infer stellar ages from rotation speeds (\citealt{Skumanich1972,Kawaler1987,Pinsonneault1989,Barnes2007, Barnes2010,Mamajek2008,Saders2013,Gallet2013,Epstein2014,Matt2015,Angus2019}, but also see \citealt{Saders2016}).

To explore how close binaries diverge from the expectations of gyrochronology, we restricted our APOGEE dwarf sample to those stars below the Kraft break, or \Teff\ $<6250$ K. We based our selection on the \Teff\ values from APOGEE rather than the masses from \cite{Sanders2018} because the former are determined directly from the spectra. The line used to select our dwarf sample in Fig.~\ref{fig:hrdiag} was chosen to retain the photometric binaries that are offset from the main sequence, but this leads to contamination from subgiants above \Teff$\gtrsim 5300$ K. We use the same line shown in Fig.~\ref{fig:hrdiag} but with a larger y-intercept to remove 1423 suspected subgiants from our sample. 

We will compare the predicted gyrochronological ages of our sample to the isochrone ages from \citet{Sanders2018}, who calculated Bayesian posteriors on masses $M$ and ages \agesd\ from fits to PARSEC isochrone using Gaia DR2 parallaxes, broadband photometry, and APOGEE spectral parameters. This method assumes single star models and introduces additional biases in some parameters, including masses and stellar ages \citep[][]{El-Badry2018a}, although our removal of SB2 systems should reduce systematic errors to some extent. In any case, stellar ages are notoriously hard to estimate without astroseismology (\citealt{Ness2016,Pinsonneault2018}, and see \citealt{Soderblom2010} for a review). While age estimates for individual stars should only be considered a starting point for a deeper analysis, general trends should be preserved in our large, statistical sample.

\begin{figure*}
	\includegraphics[width=0.85\textwidth]{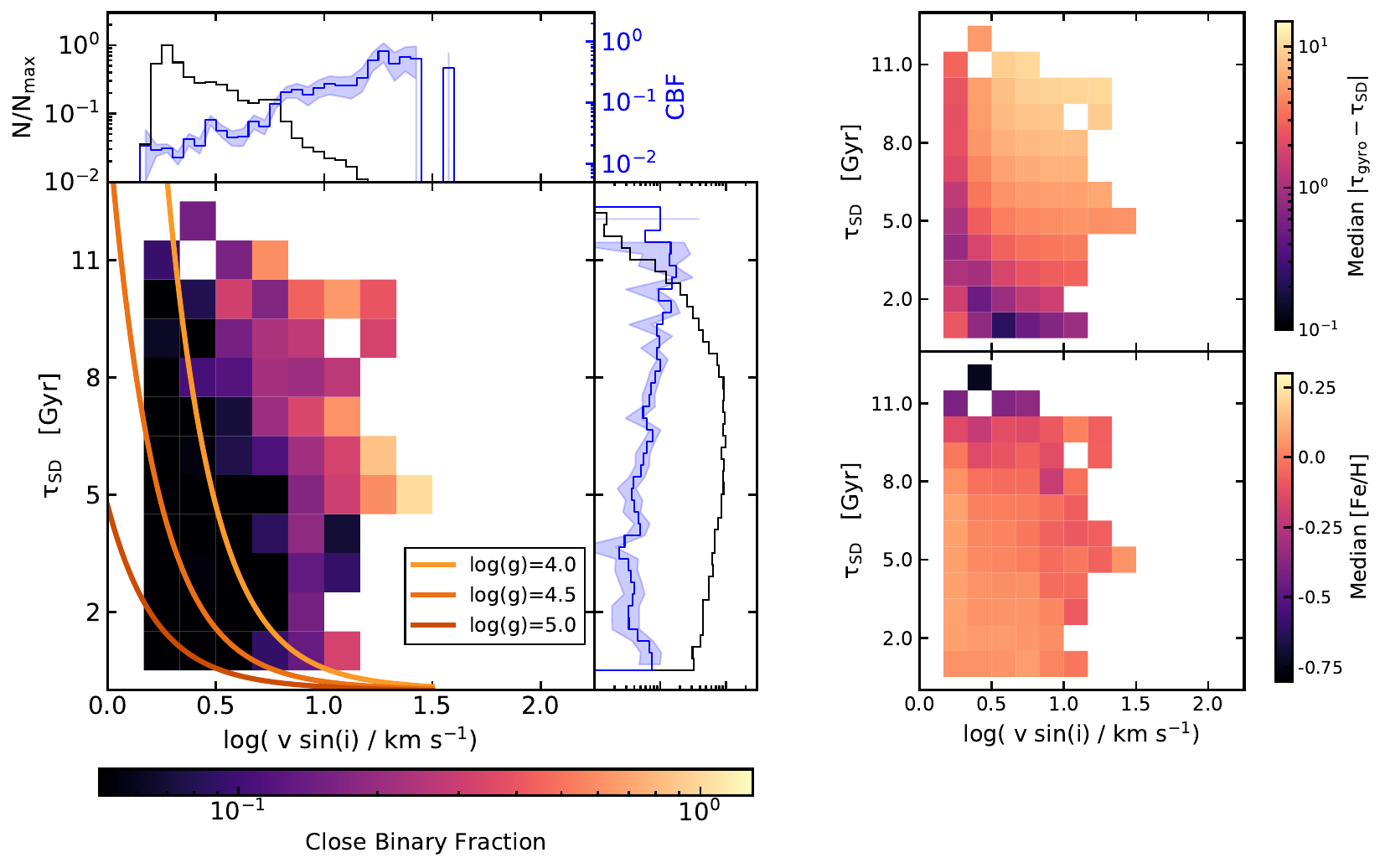}
    \caption{Left: Two-dimensional histogram showing the completeness-corrected close binary fraction as a function of stellar age $\tau_{\text{SD}}$ and rotation speed $\log($\vsini{} / \kms$)$ for dwarfs with \Teff$<6250$ K (below the Kraft break). The orange lines indicate the gyrochronology relation of \citet{Barnes2010} for 1 \msun\ and \loggunits\ $=$ 4.0 (lightest), 4.5, and 5.0 (darkest). The side panels show the normalized histogram (black) and completeness-corrected close binary fraction (blue) for each parameter alone. Blue shading indicates the uncertainties on the completeness-corrected close binary fraction. Right: The same binning scheme as the left panel, but now showing the median in the difference between predicted age and measured age, $|\tau_{\text{gyro}}-\tau_{\text{SD}}|$, and the median [Fe/H] of the stars in that bin.}
    \label{fig:kraft_break}
\end{figure*}

We plot 2D and 1D histograms of \vsini\ and stellar age \agesd\ in Fig.~\ref{fig:kraft_break}. The black histograms in the top and right panels show the distributions of each parameter, and the blue histograms represent the close binary fraction as a function of that parameter alone. In the 2D histogram on the left side of the figure, the bins are coloured by the completeness corrected close binary fraction, shown in the colorbar at the bottom. We calculate completeness corrected close binary fractions using the same procedure described in \cite{Moe2019} and \citet{Mazzola2020}, which we will briefly outline below.

 From our MC simulations, we calculated the cumulative fraction of close ($\log(P$/d$)<4$) binaries with \drvm$\geq3$ \kms, i.e., the fraction of binaries we would confidently detect. The inverse of this fraction is the factor needed to correct our sample for completeness and recover the total number of close binaries in our sample \citep[for more discussion, see Sec. 3.1 and Fig. 3 in][]{Mazzola2020}. From the MC simulation of our dwarf sample (see Section~\ref{sec:results}), we found a completeness fraction of 0.35 for \drvm\ $\geq 3$ \kms\ and $\log(\text{P}/\text{d})<4$. As discussed in \citet{Moe2019} and \citet{Mazzola2020}, we expect systematic biases that favour observing twin binaries in a magnitude-limited sample and disfavor observing SB2s with blended absorption features. Following \citet{Mazzola2020}, we assume that the Malmquist bias favouring the detection of twin binaries should be larger than the difficulties inherent in measuring SB2 RV variability, so we reduce our completeness-corrected close binary fractions by 10 per cent. Accounting for this bias leads to an estimated detection efficiency of 0.39 for close ($\log(\text{P}/\text{d})<4$) binaries in our dwarf sample when using our \drvm\ threshold.

For both the 2D and 1D histograms, we require at least ten objects per bin to cover as much parameter space as possible. We calculate the RV variability fraction $f$ (the fraction of systems with \drvm\ $\geq 3$ \kms) in each histogram bin and correct it for completeness: close binary fraction $=f/c$ where $c=0.39$ is our detection efficiency. Uncertainties in the close binary fraction are shown in the 1D histograms as shaded regions but are not shown in the 2D histogram. Uncertainties for both scale as $\sigma_{f}/c$, where $\sigma_{f}$ is the uncertainty from the binomial process on each measurement,
\begin{equation}
    \sigma_{f} = \sqrt{ \frac{f(1-f)}{N} }
\end{equation}
where $N$ is the total number of objects in each bin. As noted in \citet{Mazzola2020}, this method can result in bins with completeness-corrected close binary fractions that are larger than 100 per cent. We again assumed the same \citet{Raghavan2010} period distribution for all systems in our MC simulations, an assumption that may not be valid for the entirety of the APOGEE sample \citep{Moe2019}. Metal-poor eclipsing binaries have been found to be skewed towards shorter periods than metal-rich systems \citep{Jayasinghe2021}. A systematic shift at short periods produces an overcorrection from the completeness estimate and thus creates bins with excessively high close binary fractions. Future studies detailing the relationship between stellar chemistry and the period distribution will enable us to improve our completeness estimates.

As expected, the close binary fraction is strongly correlated with \vsini. It is unclear whether a trend exists with \agesd\ from our figure, especially given the larger uncertainties. However, age has significant internal correlations with other parameters that do correlate with the close binary fraction, such as [Fe/H] and $\alpha$ process abundances, so a thorough multi-variate analysis is necessary to fully interpret this histogram. Despite this, the 2D histogram shows a clear age-dependent gradient in the close binary fraction across \vsini, such that stars that are 10 Gyr old have notably larger fractions of close binaries at \vsini\ $<10$ \kms\ than stars that are 3 Gyr.

Due to using only single star tracks, equal-mass binaries that are offset from the main sequence are expected to be biased towards very young (100s Myr) or very old ($>10$ Gyr) \agesd\ estimates. Removing SB2s reduces the impact of these biases, but particularly for the oldest stars, we still expect some contamination from poorly-constrained binary \agesd\ in the age-dependence of the close binary fraction. However, we note that the gradient extends down into intermediate ages ($1 \lesssim$ \agesd/Gyr $\lesssim8$), where we expect relatively robust ages for binaries and single stars alike. To further check for biases from photometric binaries, we calculated a simple photometric offset $J_{\text{avg}}-J_{\text{abs}}$ from a line spanning the centre of the main sequence, such that photometric binaries should have $J_{\text{avg}}-J_{\text{abs}}\gtrsim0.5$. We coloured the 2D histogram from Fig.~\ref{fig:kraft_break} on the median photometric offset and found that stars with \agesd\ $\gtrsim 9.5$ Gyr have systematically positive offsets of $J_{\text{avg}}-J_{\text{abs}}\approx 0.5$, as to be expected for the oldest stars moving towards the subgiant branch. However, the remaining portion of the figure showed very small median offsets, with a median value of $0.05$ for histogram bins between $4.5\leq$ \agesd\ $<9.5$ Gyr, a region where we still see the effects of the age gradient.

To compare this trend with predictions from gyrochronology, we use the relation from \cite{Barnes2010},
\begin{equation}
    \tau_{\text{gyro}} = \frac{\tau_{c}}{k_{\text{C}}} \ln{\left( \frac{P_{\text{rot}}}{P_{0}} \right) } + \frac{k_{\text{I}}}{2\tau_{c}} \left(P^{2}_{\text{rot}} - P^{2}_{0} \right)
    \label{eq:gyro}
\end{equation}
where $\tau_{c}$ is the convective turnover timescale, $P_{0}$ is the initial period, and $k_{\text{C}}$ and $k_{\text{I}}$ are dimensionless constants. We adopt $\tau_{c}=34.87$ days \citep{Barnes2010a} and $P_{0}=1.1$ days \citep{Saders2016}. The orange lines in Fig.~\ref{fig:kraft_break} show this relation for 1 \msun\ and \loggunits\ $=$ 4.0 (lightest), 4.5, and 5.0 (darkest). Gyrochronology predicts that most old MS stars should be rotating slowly, but our data indicate that tidal interactions in close binary systems can keep older stars spinning faster than predicted. This manifests as a gradual increase in the close binary fraction across our measured rotation speeds as a function of stellar age.

We further explore these discrepancies with gyrochronology in the right side of the figure. We calculated \agegyro\ from equation~(\ref{eq:gyro}) with each star's APOGEE \logg, our \vsini, and \cite{Sanders2018} mass. The 2D histograms show the same binning scheme as before, but they are now coloured by the median in the difference between the predicted gyrochronological age and the measured isochrone age, $|\tau_{\text{gyro}}-\tau_{\text{SD}}|$, and the median [Fe/H] of each bin. There is clear overlap in the bins with large close binary fractions and those with large age discrepancies, some on the order of 10 Gyr or more. This provides supporting evidence for the hypothesis that the components of wide-separation binaries with unusually fast rotation rates are due to effects of close binary companions \citep{Janes2017,GodoyRivera2018}. As expected, the oldest stars have lower median [Fe/H] values, and there is a modest bias towards lower [Fe/H] among the bins with large close binary fractions, in agreement with the well-established anti-correlation between the close binary fraction and metallicty \citep[e.g.][]{Moe2019}. However, the differences in median [Fe/H] are small across much of the figure and thus unlikely to drive the observed trends. Calculating \agegyro\ with 5$P_{0}$ and $0.2P_{0}$ did not significantly change our median differences in predicted and measured ages. Repeating this analysis using the threshold \drvm\ $\geq 10$ \kms\ and detection efficiency $c=0.25$ removed five bins due to insufficient numbers of systems but revealed the same trends in all panels of Fig.~\ref{fig:kraft_break}.

 In a young field, rapid rotation is mostly a single star phenomenon, but in an old field, essentially all rapid rotators are binaries or merger products. This is evidenced by the similar fractions of rapid rotators (\vsini\ $\geq 10$ \kms) but differences in RV variability for young and old stars within our sample; young stars ($0.5\leq\tau_{\text{SD}}<3$ Gyr) have $N_{\text{fast}}/N_{\text{tot}}=0.026\pm0.003$ and old stars ($\tau_{\text{SD}}\geq8$ Gyr) have $N_{\text{fast}}/N_{\text{tot}}=0.031\pm0.003$, but Fig.~\ref{fig:kraft_break} demonstrates significant differences in the close binary fraction between the rapid rotators of these two age groups. Those wishing to apply gyrochronology relations to cool MS stars should thus be cautious and consider taking several spectra for each target to remove the RV variables, particularly for metal-poor samples, or consider using Gaia RUWE statistics to infer photcentre wobble \citep{Belokurov2020}.

\section{Conclusions}
\label{sec:concl}

 We have explored the connection between multiplicity and rotation across stellar lifetimes. Using a sample of 24 496 dwarfs and 2786 giants from APOGEE DR14, we measured trends between the maximum RV shift, \drvm, effective gravity \loggunits, and projected rotation speeds \vsini, and we interpreted these trends through the application of theoretical upper limits calculated using a simple set of assumptions. By assuming rotational synchronization and that the minimum allowed period is the critical period for Roche Lobe overflow, \pcrit, we calculated theoretical upper limits on rotational speeds and \drvm, which were able to explain the maximum extent of our data across several \logg\ bins. We simulated populations of binaries using a Monte Carlo sampler and compared their simulated \drvm\ distributions to the slow and rapid rotators in our APOGEE data. The distributions for rapid rotators in our sample were consistent with those from the shortest-period binaries in our simulation, supporting the idea that rapid rotators are more likely to possess a close companion. We also see evidence for this in dwarfs below the Kraft break: older stars show increased close binary fractions across the entire range of \vsini\ values. Older, rapidly rotating stars have particularly large close binary fractions and correspondingly large differences between their predicted gyrochronological ages and those measured by isochrone fits. Rotationally synchronized binaries present an ongoing challenge for gyrochronology, but great progress can be made to characterise these systems from the intersection of spectroscopic, photometric, and astrometric observations enabled by modern surveys of field stars.

\section*{Acknowledgements}

The authors would like to thank the anonymous referee for their constructive comments that helped to improve this manuscript. CMD and CB acknowledge support from the National Science Foundation (NSF) grant AST-1909022, and SRM acknowledges support from NSF grant AST-1616636.

Funding for the Sloan Digital Sky Survey IV has been provided by the Alfred P. Sloan Foundation, the U.S. Department of Energy Office of Science, and the Participating Institutions. SDSS acknowledges support and resources from the Center for High-Performance Computing at the University of Utah. The SDSS web site is www.sdss.org.

SDSS is managed by the Astrophysical Research Consortium for the Participating Institutions of the SDSS Collaboration including the Brazilian Participation Group, the Carnegie Institution for Science, Carnegie Mellon University, the Chilean Participation Group, the French Participation Group, Harvard-Smithsonian Center for Astrophysics, Instituto de Astrof\'{i}sica de Canarias, The Johns Hopkins University, Kavli Institute for the Physics and Mathematics of the Universe (IPMU) / University of Tokyo, the Korean Participation Group, Lawrence Berkeley National Laboratory, Leibniz Institut f\"{u}r Astrophysik Potsdam (AIP), Max-Planck-Institut f\"{u}r Astronomie (MPIA Heidelberg), Max-Planck-Institut f\"{u}r Astrophysik (MPA Garching), Max-Planck-Institut f\"{u}r Extraterrestrische Physik (MPE), National Astronomical Observatories of China, New Mexico State University, New York University, University of Notre Dame, Observat\'{o}rio Nacional / MCTI, The Ohio State University, Pennsylvania State University, Shanghai Astronomical Observatory, United Kingdom Participation Group, Universidad Nacional Aut\'{o}noma de M\'{e}xico, University of Arizona, University of Colorado Boulder, University of Oxford, University of Portsmouth, University of Utah, University of Virginia, University of Washington, University of Wisconsin, Vanderbilt University, and Yale University.

\section*{Data Availability}

APOGEE DR14 stellar parameters, abundances, and RVs are derived from the \verb+allStar+ file at \url{https://www.sdss.org/dr14/irspec/spectro\_data/}, and the \cite{Sanders2018} catalogue is downloadable from \url{https://www.ast.cam.ac.uk/jls/data/gaia_spectro.hdf5}. The likely SB2s from \cite{Mazzola2020} are available at \url{https://doi.org/10.1093/mnras/staa2859}. Rotation speeds and simulated MC data are available from C.M.D. upon reasonable request.



\bibliographystyle{mnras}
\bibliography{rotation_paper_refs_v3.1} 








\bsp	
\label{lastpage}
\end{document}